\documentclass[fleqn,11pt,letterpaper]{article}
\usepackage{amsfonts}
\usepackage{amssymb}
\usepackage{amsmath}
\usepackage{amsthm}
\usepackage{enumerate}
\usepackage{multirow}
\usepackage{float}
\usepackage{graphicx}
\usepackage{epstopdf}
\usepackage{color}
\usepackage[margin=2.57cm]{geometry}
\definecolor{darkred}{rgb}{0.5,0.2,0.2}
\usepackage[pdftitle={SRCM},pdfauthor={Mikkel Bennedsen, Eric Hillebrand, Siem Jan Koopman},colorlinks=TRUE,allcolors=darkred]{hyperref}
\usepackage{booktabs}
\usepackage{pdflscape}
\usepackage[round]{natbib}

\usepackage{tabularx}

\usepackage{float}
\floatstyle{plaintop}
\restylefloat{table}

\restylefloat{table}

\theoremstyle{plain}

\def\bi{\begin{itemize}}
\def\ei{\end{itemize}}

\allowdisplaybreaks
\graphicspath{{Figures/}}
\numberwithin{equation}{section}

\newif\ifi

\title{Matters arising:\\ Is there evidence of a trend in the CO$_2$ airborne fraction?}

\author{Mikkel Bennedsen\thanks{
Department of Economics and Business Economics and CREATES, 
Aarhus University, 
Fuglesangs All\'e 4,
8210 Aarhus V, Denmark.
E-mail:\
\href{mailto:mbennedsen@econ.au.dk}{\nolinkurl{mbennedsen@econ.au.dk}} 
},
Eric Hillebrand\thanks{
Department of Economics and Business Economics and CREATES, 
Aarhus University, 
Fuglesangs All\'e 4,
8210 Aarhus V, Denmark.
E-mail:\
\href{mailto:ehillebrand@econ.au.dk}{\nolinkurl{ehillebrand@econ.au.dk}} 
},
Siem Jan Koopman\thanks{
Department of Econometrics, 
School of Business and Economics,
Vrije Universiteit Amsterdam, 
De Boelelaan 1105,
1081 HV Amsterdam, The Netherlands,
and CREATES.
E-mail:\
\href{mailto:s.j.koopman@vu.nl}{\nolinkurl{s.j.koopman@vu.nl}} 
}}

\begin{document}
\maketitle

In a paper recently published in this journal, van Marle et al. \citep{vM2022} introduce an interesting new data set for land use and land cover change CO$_2$ emissions (LULCC)
that they use to study whether a trend is present in the airborne fraction (AF), defined as the fraction of CO$_2$ emissions remaining in the atmosphere.
Testing the hypothesis of
a trend in the AF has attracted much attention, with the overall consensus that no statistical
evidence is found for a trend in the data \citep[][]{Knorr2009,Gloor2010,Raupach2014,BHK2019a}.
In their paper, van Marle et al. analyze the AF as implied by three different LULCC emissions time series (GCP, H\&N, and their new data series).
In a Monte Carlo simulation study based on their new LULCC emissions data, van Marle et al. find evidence of a declining trend in the AF.

In this note, we
argue that the statistical analysis presented in van Marle et al. can be improved in several respects.
Specifically, the Monte Carlo study presented in van Marle et al. is not conducive to determine whether there is a trend in the AF.
Further, we re-examine the evidence for a trend in the AF by using a variety of different statistical tests.
The statistical evidence for an uninterrupted (positive or negative) trend in the airborne fraction
remains mixed at best.
When allowing for a break in the trend, there is some evidence for upward trends in both subsamples.

\section*{The Monte Carlo design of van Marle et al.}

Monte Carlo studies can be employed in situations where a test-statistic has unknown properties (size and/or power), either because it is suspected or known
that the data-generating process (DGP) does not satisfy the assumptions of the test, or because a relatively small sample size makes the invocation of a central limit theorem
for the test-statistic dubious. 
In a Monte Carlo design, the DGP needs to be well enough understood to specify a realistic and meaningful set of processes to simulate from.
To establish the size of the test, $Pr(\textrm{reject }H_0 | H_0\textrm{ is true})$, the DGP is simulated under the null hypothesis (no trend),
and the test-statistic is calculated for a large number of simulated trajectories. 
The percentiles of the ensemble of simulated test-statistics reveal how one should choose a critical value for the test in order to obtain a desired size.
One can compare the test-statistic obtained from the data to these percentiles in order to estimate the probability of observing the result from the data if the null hypothesis is true.
To determine the power of a test, $Pr(\textrm{reject }H_0 | H_1\textrm{ is true})$, the DGP is simulated under the alternative hypothesis (trend is present).
This necessitates a defendable choice of the trend specification
for the simulated DGP.
The percentiles of the ensemble of simulated test-statistics reveal how one should choose a critical value for the test in order to obtain a desired power.

In their paper, Marle et al. perturb the original data with simulated Gaussian random variables, where variances are chosen that correspond
to empirical variances of the data time series.
The non-parametric Mann-Kendall statistic is then computed to test for a trend in 
the simulated AF time series. These two steps are repeated $10,000$ times, and the number of times a positive or
negative trend is detected by the Mann-Kendall test on the simulated data is counted (Table 1 in van Marle et al.).
While this design has some elements of the procedure outlined above, it falls short of giving the researcher a handle on whether they are simulating under
the null hypothesis of no trend or under the alternative hypothesis.
Therefore, the frequency of detected trends in the design of van Marle et al. cannot be interpreted as a probability under either hypothesis. Similarly, it cannot be interpreted as a probability that the original data contain a trend, 
 as appears to be suggested in van Marle et al.

\section*{How much statistical evidence is there for a negative trend?}

There are many different ways to test for the presence of a trend, just as there are many conceptualizations of trends \citep{hamilton1994}.
In the Monte Carlo simulation study of van Marle et al.,
the non-parametric Mann-Kendall statistic is used to test for a trend.
However, the Mann-Kendall test-statistic is not computed for the actual data set.
When applied to the data, no statistical evidence of a trend is found (Table \ref{tab:tab1_MK_main}). 

\begin{table}
\caption{\it $p$-values for various tests on the AF data from van Marle et al. The $p$-values are reported for the Mann-Kendall tests (
$H_0^{MK}: \textnormal{``data do not contain a trend''}$ against the two-sided alternative $H_1: \textnormal{``trend in data''}$;
$H_1^{MK+}$: the one-sided alternative for ``positive trend in data'';
$H_1^{MK-}$: the one-sided alternative for ``negative trend in data''), for the trend slope coefficients (
$H_0^{\textrm{slope1}}: b = 0$ against the two-sided alternative $H_1: b \neq 0$ for the linear trend model, $AF_t = a + bt + \epsilon_t$;
$H_0^{\textrm{slope2}}: b = 0$ against the two-sided alternative $H_1: b \neq 0$ for the linear trend model with intercept break in 1988 for raw series and in 1990 for filtered series,  $AF_t = a_1 + bt + a_2 I(t \geq \tau)+ \epsilon_t$) and for the break coefficients (
$H_0^{\textrm{br-trend}}: b_2 = 0$ against the two-sided alternative $H_1: b_2 \neq 0$ for the linear trend model with break, $AF_t = a_1  +  b_1  t + a_2 I(t \geq \tau) +  b_2(t-\tau)I(t\geq \tau) + \epsilon_t$;
$H_0^{\textrm{br-intercept}}: a_2 = 0$ against the two-sided alternative $H_1: a_2 \neq 0$ for the linear trend model with break, $AF_t = a_1  +  b  t + a_2 I(t \geq \tau) +   \epsilon_t$),
%
%
for the three different data series studied in van Marle et al. (GCP, H$\&$N, New), using both the raw and filtered versions (`raw' and `filter').}
\label{tab:tab1_MK_main}
\begin{center}
\scriptsize
\begin{tabular}{llcccccccc@{}} 
\toprule
& & \multicolumn{3}{c}{Mann-Kendall tests} & & \multicolumn{4}{c}{Regression tests}\\
& & $H_0^{MK}$  & $H_0^{MK+}$  & $H_0^{MK-}$ & & $H_0^{\textrm{slope1}}$ & $H_0^{\textrm{slope2}}$  & $H_0^{\textrm{br-trend}}$ & $H_0^{\textrm{br-intercept}}$ \\  
\midrule
GCP  & raw       & 0.1017  &   0.0508 & 0.9492& & 0.1447 & 0.0174 & 0.1791  & 0.0456   \\
     & filter    & 0.0646  &   0.0323 & 0.9677 & & 0.0634 & 0.0030 &   0.2254  & 0.0213   \\
H\&N & raw       & 0.4819  &   0.2409 & 0.7591 & & 0.8262 & 0.0184 &    0.3514    &  0.0096   \\
     & filter    & 0.6230  &   0.3115 &  0.6885 & & 0.8738 & 0.0048 &  0.7698  & 0.0017   \\
New  & raw       & 0.7415  & 0.6292 & 0.3708 &   & 0.3473 & 0.0184 &   0.3692   & 0.0096   \\
     & filter    & 0.1850  & 0.9075 & 0.0925 &   & 0.1130 & 0.0474 &   0.8109 & 0.0016   \\
\bottomrule 
\end{tabular}
\end{center}

\end{table}

\begin{figure}[h!] 
\centering 
\includegraphics[width=0.95\columnwidth]{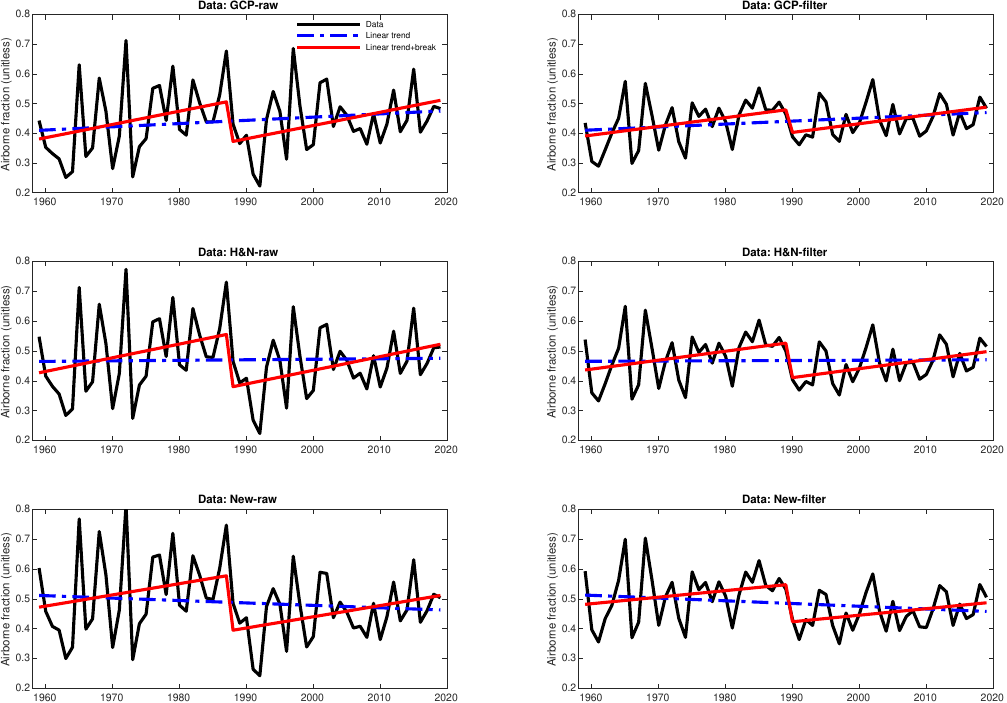} 
\caption{\it AF data (black, solid), estimated trends from the linear trend model (blue, dashed), and the linear trend model with a break (red, solid).}
\label{fig:trbr0}
\end{figure}

To complement the Mann-Kendall test, we perform a number of further tests for a trend in the AF, none of them finding evidence of a declining trend in the AF (Supplementary Information). Here, we report the results for 
the test that most researchers would resort to at first pass: the $t$ test for for the least-squares estimate of a linear trend in the linear model $AF_t = a + b t + \epsilon_t$.
The estimated slope coefficient $\hat b$ is positive
for the GCP and H\&N data series, while for the AF series computed from the new LULCC series, the slope estimate is indeed negative (Supplementary Information, Table \ref{tab:trbr00}).
These statements hold for both the raw and the filtered series.
Figure \ref{fig:trbr0} presents the fitted trends (blue dashed line).
The corresponding tests indicate that the slope estimates are not significant for any of the AF time series 
($H_0^{\textrm{slope1}}$, Table \ref{tab:tab1_MK_main}).

Breaks in the level and trend of the AF have been discussed in Keenan et al. \citep{keenan2016recent}.
The time series econometrics literature has a tradition of studying these ``structural breaks'' \citep{perron2006,clements2006,mills2010skinning}.
We consider the single-break model \citep{perron2005structural}:
\begin{equation*}
AF_t = a_1  +  b_1  t + a_2 I(t \geq \tau) +  b_2 (t-\tau)I(t\geq \tau) + \epsilon_t, 
\end{equation*}
where $\tau \geq 1$ and $\epsilon_t$ is a covariance-stationary error. This model allows for a change in the intercept from $a_1$ to $a_1 + a_2$ and a change in the trend slope from $b_1$ to $b_1 + b_2$ at time $\tau$.
The coefficients are estimated for each $\tau=11,\ldots,51$ (leaving out 10 observations at each side of the sample to avoid degeneracies).
The break point $\tau$ is chosen where the sum of squared errors is lowest. 

This method finds break points in the three raw series in 1988 and in the three filtered series in 1990.
It is not without peril to connect statistically found change points to historical events.
In the case of the AF data, relevant events may be the Pinatubo minimum in the change in atmospheric concentrations \citep{bousquet2000regional,angert2004co2}, the collapse of the Soviet Union and the first Gulf war in anthropogenic emissions, and fire events documented for this period in the LULCC series \citep{vM2022}.
We can subsequently test for whether the data support a break in the trend, by testing the hypothesis $b_2 = 0$. For all data series, 
this hypothesis cannot be rejected at a nominal level of $5\%$ ($H_0^{\textrm{br-trend}}$, Table \ref{tab:tab1_MK_main}), indicating that there is no break in the trend. 
Conversely,  the hypothesis $a_2 = 0$ can be rejected at a nominal level of $5\%$ ($H_0^{\textrm{br-intercept}}$, Table \ref{tab:tab1_MK_main}), indicating that the break is driven by a change in the intercept.
Figure \ref{fig:trbr0} presents the resulting estimated trends (red solid line).
All segmented trend estimates are positive (Supplementary Information, Table \ref{tab:trbr}) and highly significant ($H_0^{\textrm{slope2}}$, Table \ref{tab:tab1_MK_main}). 
%
%

\clearpage

\newpage

 {\small 
\bibliographystyle{chicago}
\bibliography{bhk_references}
}

\clearpage

\newpage

\appendix

\section{Supplementary Information}\label{app:OLS}

\subsection{Evidence from linear trend regressions, with and without breaks}
We consider the linear trend regression model
\begin{align}\label{eq:OLS}
AF_t = a + b t + \epsilon_t,
\end{align}
where $\epsilon_t$ is an error term and $AF_t$ is one of the measures of AF (e.g. the `raw' or `filtered' AF).
The purpose is to statistically test the null hypothesis of ``no trend'', formally given as
\begin{align*}
H_0: b = 0,
\end{align*}
against one of the following alternative hypotheses
\begin{align*}
H_0^{two-sided}: b \neq 0, \\
H_0^{pos}: b>0, \\
H_0^{neg}: b <0.
\end{align*}
We can perform tests on the parameter $b$ using the least squares method. We consider the $t$ test-statistic
\begin{align*}
t = \frac{\hat b}{s.e(\hat b)},
\end{align*}
where $\hat b$ is the least squares estimator of $b$ from Equation \eqref{eq:OLS} and $s.e(\hat b)$ is the standard error of this estimator,
which we have calculated using the heteroskedasticity and autocorrelation robust (HAC) estimator of $s.e.(\hat b)$ proposed by \cite{newey1987simple}.
We also present the results obtained from using the standard estimator of $s.e(\hat b)$ which lead to similar findings and conclusions.

Table \ref{tab:trbr00} presents the regression estimation results obtained from the least squares method applied to Equation \eqref{eq:OLS}, while
Table \ref{tab:tab1_OLS2} presents the $p$-values from the test of $H_0: b=0$.
From the results in Table \ref{tab:tab1_OLS2} we learn that it is not possible to conclude that there is a trend in the data (either positive or negative),
no matter which data set is used.

\begin{table}[h!]
\caption{\footnotesize \it \textbf{Least-squares regression output of the linear model.} Output of the least-squares regression of the model $y_t = a + b t  + \epsilon_t$, where $t = 0, 1, \ldots, 60$. Standard errors are calculated using the \cite{newey1987simple} HAC estimator.}
\vspace{4mm}
\centering
\scriptsize
\begin{tabular}{lcccccc@{}} \hline\hline
Data: &  GCP-raw  &  GCP-filter &  H\&N-raw  &  H\&N-filter &  New-raw  &  New-filter \\ \hline 
$\hat a$ (intercept)    &     0.4108   &  0.4113   &  0.4657   &  0.4662  &   0.5120  &   0.5132  \\
SE($\hat a$)   &    0.0308  &   0.0213  &   0.0340  &   0.0211  &   0.0363  &   0.0226   \\
$t$-stat &      13.3436 &   19.3437 &   13.7112  &  22.1169  &  14.1083 &   22.6800    \\
\cmidrule{2-7}
$\hat b$ (trend)    & 0.0011&     0.0010  &   0.0002  &   0.0001 &   -0.0008  &  -0.0009   \\
SE($\hat b$)   &       0.0007  &   0.0005   &  0.0008  &   0.0005  &   0.0009 &    0.0006\\
$t$-stat &      1.4585  &   1.8567   & 0.2195  &   0.1589  &  -0.9398 &   -1.5847\\  \hline
\end{tabular}
\label{tab:trbr00}
\end{table}

%

\begin{table}[h!]
\caption{\footnotesize \it \textbf{Testing for a trend in AF using least squares and HAC standard errors.} Testing $H_0: b = 0$ using least squares and HAC standard errors. We have carried out the test for the three different data series studied in the main paper (GCP, H\&N, New), as well as using the raw and filtered data (`raw' and `filter').}
\vspace{4mm}
\centering
\scriptsize
\begin{tabular}{lcccccc@{}} \hline\hline
  & $p$-value ($H_0^{two-sided}$)  & $p$-value ($H_0^{pos}$)  & $p$-value ($H_0^{neg}$)  \\\hline  
GCP (raw)              &   0.1447 &   0.0724    &0.9276  \\
GCP (filter)              &   0.0634  &  0.0317   & 0.9683  \\
H\&N (raw)              &   0.8262   & 0.4131 &   0.5869    \\
H\&N (filter)             &     0.8738  &  0.4369  &  0.5631      \\
New  (raw)              &   0.3473  &  0.8263  &  0.1737  \\
New  (filter)              &  0.1130  &  0.9435  &  0.0565  \\ \hline
\end{tabular}
\label{tab:tab1_OLS2}
\end{table}


Next, the focus is on the statistical evidence of a trend in AF, after allowing for a break in the level and in the slope of AF.
We consider the single-break model \citep{perron2005structural}:
\begin{align}\label{eq:breq}
AF_t = a_1 + a_2 I(t \geq \tau) + b_1 t + b_2 t I(t\geq \tau ) + \epsilon_t, 
\end{align}
where $\epsilon_t$ is a covariance-stationary error. This model allows for a change in the intercept from $a_1$ to $a_1 + a_2 $ and a change in the
trend slope from $b_1$ to $b_1 + b_2$ at time $\tau$.
The coefficients are estimated for each $\tau=11,\ldots,51$ (leaving out 10 observations at each side of the sample to avoid degeneracies).
The break point $\tau$ is chosen where the sum of squared errors is lowest.
This method detects single-break points in the three raw series in 1988 and in the three filtered series in 1990.

Table \ref{tab:trbr3} presents the estimation results obtained from the least squares method applied to Equation \eqref{eq:breq};
we find that the estimates of $b_2$ are not significantly different from zero.
Hence, we conclude that the data prefer a model with no break in the slope of the trend.

\begin{table}[h!]
\caption{\footnotesize \it \textbf{Least-squares regression output of the linear model with break in both intercept and trend.} Output of the least-squares regression of the model $AF_t = a _1+ b_1 t + a_2 \cdot I(t \geq \tau) + b_2 \cdot (t - \tau) \cdot I(t \geq \tau)  + \epsilon_t$, where $t = 0, 1, \ldots, 60$ and $\tau = 30$ for the raw data series (corresponding to a break in year $1988$) and $\tau = 32$ for the filtered data series (corresponding to a break in year $1990$). Standard errors are calculated using the \cite{newey1987simple} HAC estimator.}
\vspace{4mm}
\centering
\scriptsize
\begin{tabular}{lcccccc@{}} \hline\hline
  Data: &  GCP-raw  &  GCP-filter &  H\&N-raw  &  H\&N-filter &  New-raw  &  New-filter \\ \hline 
$\hat a_1$ (intercept)    &    0.3477   & 0.3755   & 0.4029&    0.4327 &   0.4478&    0.4785  \\
SE($\hat a_1$)   &    0.0390   & 0.0228    &0.0411 &   0.0248 &   0.0425  &  0.0263  \\
$t$-stat &  8.9245  & 16.4775   & 9.7948  & 17.4312  & 10.5381  & 18.1986 \\
\cmidrule{2-7}
$\hat b_1$ (trend)    &  0.0069 &   0.0040&    0.0064 &   0.0033 &   0.0055 &   0.0024  \\
SE($\hat b_1$)   &       0.0022   & 0.0014  &  0.0025 &   0.0016  &  0.0026 &   0.0017\\
$t$-stat &     3.1870 &   2.9582   & 2.5457  &  2.0827  &  2.1010  &  1.3977\\
\cmidrule{2-7}
$\hat a_2$ (break, intercept)   &    -0.1462  & -0.0791  & -0.1856   &-0.1185&   -0.1921 &  -0.1262\\
SE($\hat a_2$)   &     0.0516  &  0.0288  &  0.0533 &   0.0299   & 0.0536 &   0.0313\\
$t$-stat &    -2.8324 &  -2.7444  & -3.4826  & -3.9649 &  -3.5827 &  -4.0324 \\
\cmidrule{2-7}
$\hat b_2$ (break, trend)   &  -0.0042&   -0.0023  & -0.0031  & -0.0006  & -0.0031 &  -0.0005\\
SE($\hat b_2$)   &    0.0032  &  0.0019   & 0.0033   & 0.0021  &  0.0034  &  0.0022\\
$t$-stat &  -1.3434  & -1.2124  & -0.9318  & -0.2927 &  -0.8980  & -0.2392 \\ \hline
\end{tabular}
\label{tab:trbr3}
\end{table}

These findings provide some support in considering the model
\begin{align}\label{eq:breq3}
y_t = a_1 + b t + a_2 \cdot I(t \geq \tau) + \epsilon_t,
\end{align} 
where $t = 0, 1, \ldots, 60$ and $\tau = 30$ for the `raw' data series (corresponding to a break in year $1988$) and
$\tau = 32$ for the `raw' data series (corresponding to a break in year $1990$). Table \ref{tab:trbr} presents the estimation results obtained from the least squares method applied
to Equation \eqref{eq:breq3}.

\begin{table}[h!]
\caption{\footnotesize \it \textbf{Least-squares regression output of the linear model with break in intercept.} Output of the least-squares regression of the model $AF_t = a_1 + b t + a_2 \cdot I(t \geq \tau) + \epsilon_t$, where $t = 0, 1, \ldots, 60$ and $\tau = 30$ for the `raw' data series (corresponding to a break in year $1988$) and $\tau = 32$ for the `raw' data series (corresponding to a break in year $1990$). Standard errors are calculated using the \cite{newey1987simple} HAC estimator.}
\vspace{4mm}
\centering
\scriptsize
\begin{tabular}{lcccccc@{}} \hline\hline
Data: &  GCP-raw  &  GCP-filter &  H\&N-raw  &  H\&N-filter &  New-raw  &  New-filter \\ \hline 
$\hat a$ (intercept)    &  0.3817   & 0.3918  &  0.4278  &  0.4370  &  0.4726  &  0.4822   \\
SE($\hat a_1$)   &   0.0298  &  0.0184   & 0.0312   & 0.0198 &   0.0322  &  0.0210   \\
$t$-stat &    12.8039  & 21.2541  & 13.7012  & 22.0220 &  14.6593  & 22.9481    \\
\cmidrule{2-7}
$\hat b$ (trend)    &  0.0045  &  0.0029   & 0.0046  &  0.0030  &  0.0038  &  0.0022   \\
SE($\hat b$)   &       0.0014   & 0.0009  &  0.0014  &  0.0010  &  0.0015 &   0.0011\\
$t$-stat &     3.2914  &  3.0982  &  3.1824  &  3.0596 &   2.5415  &  2.0714\\
\cmidrule{2-7}
$\hat a_2$ (break, intercept)   &  -0.1378&   -0.0791 &  -0.1794  & -0.1185  & -0.1860  & -0.1262\\
SE($\hat a_2$)   &    0.0519    &0.0301  &  0.0529  &  0.0298   & 0.0532   & 0.0312\\
$t$-stat &  -2.6531  & -2.6280  & -3.3905 &  -3.9760 &  -3.4993 &  -4.0474 \\ \hline
\end{tabular}
\label{tab:trbr}
\end{table}

%

For completeness, we also compare the two models in \eqref{eq:breq} and \eqref{eq:breq3} using statistical goodness-of-fit tests.
We compare the likelihoods of the two models, evaluated at their optimized parameters and assuming that the error term $\epsilon_t$ is
independent, identically and normally distributed, by means of the 
Bayesian Information Criteria (BIC). Since the model in \eqref{eq:breq} nests the model in \eqref{eq:breq3}, it is necessarily the case that the former
has a higher likelihood than the latter. When looking at the BIC, however, we see that the smaller model  \eqref{eq:breq3} has a lower BIC value than the
larger model  \eqref{eq:breq}, implying that the data prefer the smaller model.
Finally, since the two models are nested, we can also consider a likelihood ratio test and test the hypothesis $H_0: b_2=0$ in Equation \eqref{eq:breq}.
Given the results presented in Table \ref{tab:comp}, this null hypothesis cannot be rejected.
These results confirm the finding above: the data prefer a model with a break in the intercept, but no break in the trend.

\begin{table}[h!]
\caption{\footnotesize \it \textbf{Statistical comparison of the two models.} Statistical comparison of the model with break in intercept (Model 1) and the model with break in intercept and trend (Model 2). Model 1:  $AF_t = a_1 + b t + a_2 \cdot I(t \geq \tau) + \epsilon_t$. Model 2):  $AF_t = a_1 + b_1 t + a_2 \cdot I(t \geq \tau) + b_2 \cdot (t - \tau) \cdot I(t \geq \tau)  + \epsilon_t$. Here, $t = 0, 1, \ldots, 60$ and $\tau = 30$ for the `raw' data series (corresponding to a break in year $1988$) and $\tau = 32$ for the `filter' data series (corresponding to a break in year $1990$). ``logL'' is the log-likelihood of the model obtained by least-squares regression, i.e. $logL = -\frac{1}{2} n \log (SSE/n)$, where $SSE$ is the sum of squared errors and $n = 61$ is the number of observations. ``BIC'' is the Bayesian information criterion given by $BIC = -2logL + k \log n$, where $k$ is the number of parameters in a given model, so that $k = 3$ in the case of Model 1 and $k = 4$ in the case of Model 2. ``LR-stat'' is the likelihood ratio test-statistic for the null that the break in the trend is absent, i.e. $b_2 = 0$, that is $LR = -2(logL_1 - logL_2)$. Under the null, we have $LR \sim \chi^2(1)$, meaning that the critical value for the likelihood ratio (LR) test at a $5\%$ level is  $3.8415$. }
\vspace{4mm}
\centering
\scriptsize
\begin{tabular}{lcccccc@{}} \hline\hline
Data: &  GCP-raw  &  GCP-filter &  H\&N-raw  &  H\&N-filter &  New-raw  &  New-filter \\  \hline
logL (Model 1) &  137.0926 & 168.0499  &134.2751&  163.5550 & 132.3149 & 160.0704\\  
logL (Model 2) & 138.0433  &168.8265  &134.7362&  163.6008&  132.7434  &160.1010\\  
\cmidrule{2-7}
BIC (Model 1) &   -261.8527 &-323.7672 &-256.2176 &-314.7773 &-252.2972 &-307.8082\\  
BIC (Model 2) &  -259.6432& -321.2095 &-253.0289& -310.7581& -249.0433 &-303.7586\\  
\cmidrule{2-7}
LR-stat & 1.9014  &  1.5531 &   0.9222 &   0.0916   & 0.8570  &  0.0612\\  \hline 
\end{tabular}
\label{tab:comp}
\end{table}

\subsection{Evidence from year-to-year changes}

\begin{figure}[h!]
    \centering
    \caption{Running average change with increasing sample of year-to-year changes for the three different data series studied in the main paper (row-wise: GCP, H\&N, New), as well as using the raw and filtered data (column-wise: `raw' and `filter').
    Running average is defined as $(\tau -1) ^{-1}\sum _{t=2}^{\tau} \Delta y_t$, for $\tau = 2,\ldots,T$.}
    \includegraphics[width=0.9\textwidth]{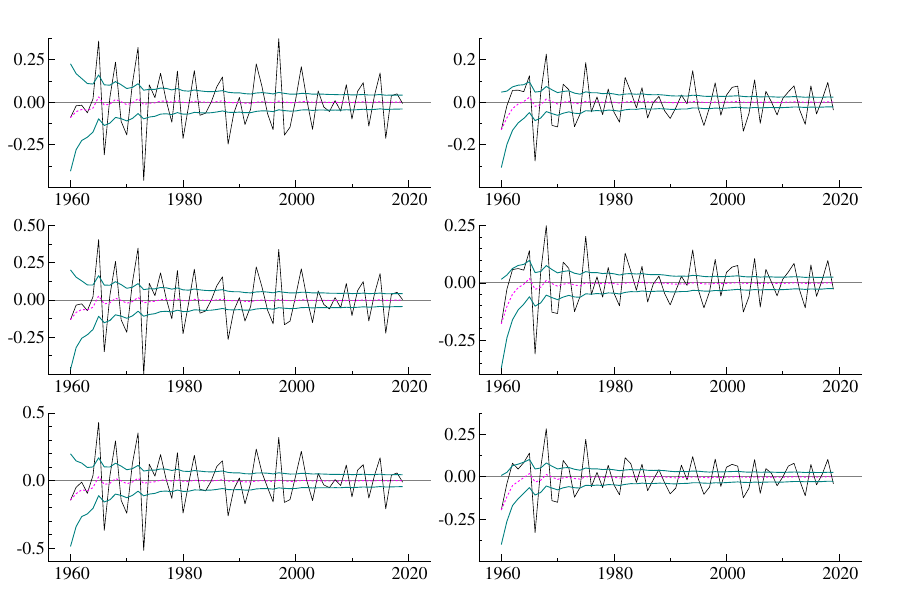}
    \label{fig:movavggr}
\end{figure}

Year-to-year changes in a time series $y_t$ are measured as $\Delta y_{t} = y_{t} - y_{t-1}$.
The average change
\[
\frac{1}{T-1} \sum _{t=2}^T \Delta y_{t},
\]
is an indication of the direction of a trend, if there is one: a negative average indicates a negative trend.
We consider the running average change defined as $(\tau-1) ^{-1}\sum _{t=2}^{\tau} \Delta y_t$, for $\tau = 2,\ldots,T$,
together with the 95\% point-wise confidence intervals for each change average (Figure \ref{fig:movavggr}).
All presented averages indicate change estimates which are not significantly different from zero.

\subsection{Evidence from stochastic trends}
\begin{figure}[h!]
    \centering
    \caption{Level $\mu _t$ estimates, with 95\% confidence interval, from the local level model for the three different data series studied in the main paper (row-wise: GCP, H\&N, New), as well as using the raw and filtered data (column-wise: `raw' and `filter').}
    \includegraphics[width=0.9\textwidth]{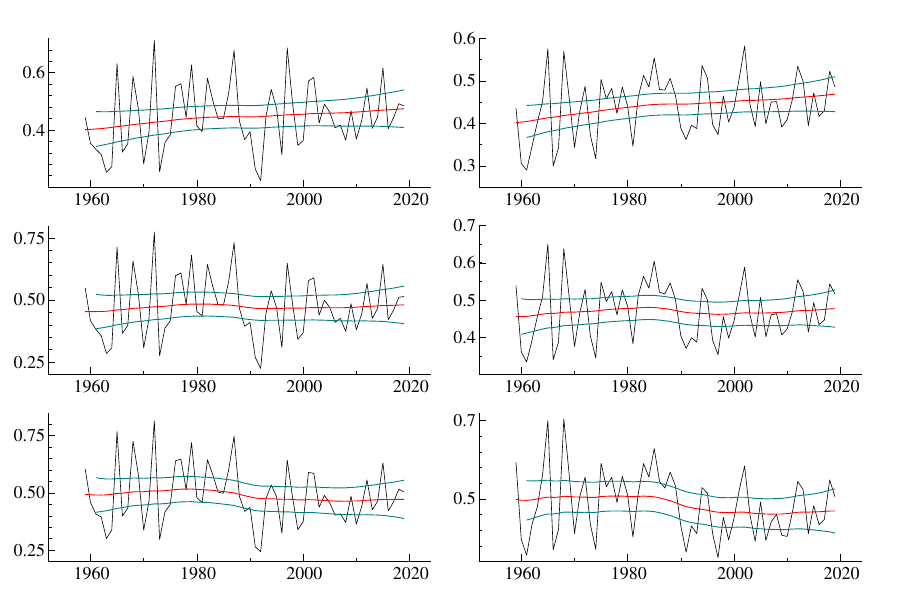}
    \label{fig:locallevel}
\end{figure}

Trends can be conceptualized as deterministic or as stochastic \citep{johansen1995,fuller1996}. A standard model for the simplest form of a stochastic trend, a random walk, is the local level model \citep{harvey1989,durbin2012time}. The local level model is also known as the random walk plus noise model, and it is given by
\[
y_t = \mu _t + \varepsilon _t, \qquad \mu _{t+1} = \mu _t + \eta _t, \qquad t=1,\ldots, T,
\]
where $\mu _1$ is treated as unknown. The estimates of the level $\mu _t$ (the stochastic trend) using all observations $y_1,\ldots,y_T$ are presented in Figure \ref{fig:locallevel}. The 95\% point-wise confidence intervals are also provided. 

\begin{figure}[h!]
    \centering
    \caption{$t$ tests for break at time $\tau$ in local level estimates, with $\tau =1,\ldots,T$, for the three different data series studied in the main paper (row-wise: GCP, H\&N, New), as well as using the raw and filtered data (column-wise: `raw' and `filter').}
    \includegraphics[width=0.9\textwidth]{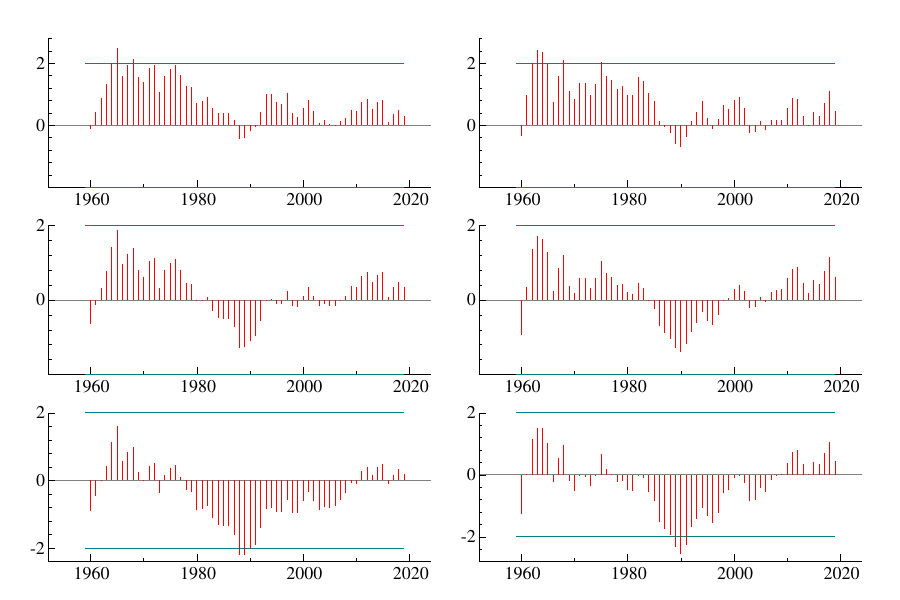}
    \label{fig:breaklevel}
\end{figure}

The evidence for trends in all series is rather weak. The estimated stochastic trends for the GCP and H\&N data series (raw and filtered) show slowly upward moving trends.
The new AF series have estimated trends that are somewhat downward moving. However, the patterns are not showing continuous declines.
These trends reveal that they are more or less constant from 1959 to about 1990, then shifts take place to lower levels, which remain more or less constant, that is, 
from the mid-90s to the end of the sample. To follow-up on these findings,
we introduce a break dummy $I_{\{t>\tau\}}$ in the random walk specification for the trend $\mu _t$,
which takes value one for $t>\tau$ and zero elsewhere.
We record the $t$ tests of the break dummy $I_{\{t>\tau\}}$ for each $\tau = 1,\ldots,T$ in
Figure \ref{fig:breaklevel}. There is little evidence of a break in the GCP and H\&N time series.
There is, however, strong evidence of a break in the AF series (raw and filtered) around the years 1988-1991.
These findings confirm our earlier conclusions.


\clearpage

\end{document}